\documentstyle[aps,multicol,pra,amsfonts]{revtex}
\newtheorem{lemma}{Lemma}
\newtheorem{thrm}{Theorem}
\newtheorem{cor}{Corollary}
\newtheorem{ex}{Example}
\tighten
\begin{document}
\draft
\title{Quantum Convolutional Error Correcting Codes}
\author{H. F. Chau\footnote{e-mail: hfchau@hkusua.hku.hk}}
\address{Department of Physics, University of Hong Kong, Pokfulam Road, Hong
 Kong}
\date{\today}
\preprint{HKUPHYS-HFC-11; quant-ph/9712029}
\maketitle
\begin{abstract}
 I report two general methods to construct quantum convolutional codes for
 $N$-state quantum systems. Using these general methods, I construct a quantum
 convolutional code of rate 1/4, which can correct one quantum error for every
 eight consecutive quantum registers.
\end{abstract}
\medskip
 A quantum computer is more efficient than a classical computer in useful
 applications such as integer factorization \cite{Factorization} and database
 search \cite{Database}. However, decoherence remains one of the major
 obstacles to building a quantum computer \cite{Decoherence}. Nevertheless, the
 effect of decoherence can be compensated for if one introduces redundancy in
 the quantum state. By first encoding a quantum state into a larger Hilbert
 space $H$. Then by projecting the wave function into a suitable subspace $C$
 of $H$. And finally by applying a unitary transformation to the orthogonal
 complement of $C$ according to the measurement result; it is possible to
 correct quantum errors due to decoherence. This scheme is called the quantum
 error correction code (QECC) \cite{9-bit}. Many QECCs have been discovered
 (see, for example,
 Refs.~\cite{9-bit,5-bit,Cald1,St1,St2,Cald2,Gott,9-reg,5-reg,N-add,AnE1,AnE2})
 and various theories on the QECC have also been developed (see, for example,
 Refs.~\cite{St2,Cald2,Gott,9-reg,5-reg,N-add,AnE1,AnE2,Cond1,Cond2,Cond3}). In
 particular, the necessary and sufficient condition for a QECC is
 \cite{Cond1,Cond2,Cond3,Gottesman_Thesis}
\begin{equation}
 \langle i_{\rm encode} | {\cal A}^{\dag} {\cal B} | j_{\rm encode} \rangle =
 \Lambda_{{\cal A},{\cal B}} \,\delta_{ij} ~, \label{E:Condition}
\end{equation}
 where $|i_{\rm encode} \rangle$ denotes the encoded quantum state $|i\rangle$
 using the QECC; ${\cal A},{\cal B}$ are the possible errors the QECC can
 handle; and $\Lambda_{{\cal A},{\cal B}}$ is a complex constant independent of
 $|i_{\rm encode} \rangle$ and $|j_{\rm encode} \rangle$. Note that the above
 condition for a QECC is completely general, working for finite or infinite
 number of $N$-state quantum registers.\footnote{Perhaps the simplest way to
 see that Eq.~(\protect\ref{E:Condition}) holds for infinite number of
 $N$-state registers is to observe that Gottesman's proof in
 Ref.~\protect\cite{Gottesman_Thesis} does not depend on the finiteness of the
 Hilbert space for encoded state.}
\par
 All QECCs discovered so far are block codes. That is, the original state ket
 is first divided into {\em finite} blocks of the same length. Each block is
 then encoded separately using a code which is {\em independent} of the state
 of the other blocks ({\em cf.} Refs.~\cite{Convolution1,Convolution2}).
\par
 In addition to block codes, convolutional codes are well known in classical
 error correction. Unlike a block code, the encoding operation depends on
 current as well as a number of past information bits
 \cite{Convolution1,Convolution2}. For instance, given a (possibly infinite)
 sequence of classical binary numbers $( a_1,a_2,\ldots ,a_m,\ldots )$, the
 encoding $( b_1,c_1,b_2,c_2,\ldots ,b_m,c_m,\ldots )$ with
\begin{equation}
 b_i = a_i + a_{i-2} \bmod 2, ~c_i = a_i + a_{i-1} + a_{i-2} \bmod 2
 \label{E:Classical_Conv}
\end{equation}
 for all $i$, and $a_{0} = a_{-1} = 0$ is able to correct up to one error for
 every two consecutive bits \cite{21-encoder}.
\par
 In classical error correction, good convolutional codes often can encode with
 higher efficiencies than their corresponding block codes in a noisy channel
 \cite{Convolution1,Convolution2}. It is, therefore, instructive to find
 quantum convolutional codes (QCC) and to analyze their performance. In this
 letter, I first report a way to construct a QCC from a known quantum block
 code (QBC). Then I discuss a way to construct a QCC from a known classical
 convolutional code. Finally, I report the construction of a QCC of rate 1/4
 which can correct one quantum error for every eight consecutive quantum
 registers.
\par
 Let me first introduce some notations before I construct QCCs. Suppose each
 quantum register has $N$ orthogonal eigenstates for $N\geq 2$. Then, the basis
 of a general quantum state consisting of many quantum registers can be written
 as $\left\{ |{\bf k} \rangle \right\} \equiv \left\{ |k_1,k_2,\ldots ,k_m,
 \ldots\rangle \right\}$ for all $k_m \in {\Bbb Z}_N$. And I abuse the notation
 by defining $k_m = 0$ for all $m\leq 0$.
\par
 Suppose $|k\rangle \longmapsto \sum_{i_1,i_2,\ldots ,i_m} a^{(k)}_{i_1,i_2,
 \ldots ,i_m} \,|i_1,i_2,\ldots ,i_m\rangle$ be a QBC mapping one quantum
 register to a code of length $m$. Hence, the rate of the code equals $1/m$.
 The effect of decoherence can be regarded as an error operator acting on
 certain quantum registers. I denote the set of all possible errors that can be
 corrected by the above quantum block code by $E$. Based on this QBC, one can
 construct a family of QCCs as follows:
\begin{thrm}
 Given the above QBC and a quantum state $|{\bf k}\rangle \equiv |k_1,k_2,
 \ldots ,k_n,\ldots\rangle$ making up of possibly infinitely many quantum
 registers, then the encoding
 \begin{eqnarray}
  & & |{\bf k}\rangle \equiv |k_1,k_2,\ldots ,k_n,\ldots\rangle \nonumber \\ &
  \longmapsto & |{\bf k}_{\rm encode}\rangle \equiv \bigotimes_{i=1}^{+\infty}
  \left[ \,\sum_{j_{i1},j_{i2},\ldots ,j_{im}} \!\!\!\! a^{(\sum_p \mu_{ip} k_p
  )}_{j_{i1},j_{i2},\ldots ,j_{im}} |j_{i1},j_{i2},\ldots ,j_{im}\rangle
  \right] \,, \label{E:Encode_QBC}
 \end{eqnarray}
 forms a QCC of rate $1/m$ provided that the matrix $\mu_{ip}$ is invertible.
 This QCC can handle errors in the form $E\otimes E\otimes \cdots$.
 \label{Thrm:QBC}
\end{thrm}
\noindent
{\it Proof:} I consider the effects of errors ${\cal E} \equiv {\cal E}_1
 \otimes {\cal E}_2 \otimes \cdots$ and ${\cal E}' \equiv {\cal E}_1' \otimes
 {\cal E}_2' \otimes \cdots \in E \otimes E\otimes \cdots$ on the encoded
 quantum registers by computing
\begin{eqnarray}
 & & \langle {\bf k}_{\rm encode}' | {\cal E}'^{\dag} {\cal E} |
 {\bf k}_{\rm encode} \rangle \nonumber \\ & = & \prod_{i=1}^\infty \left[ \,
 \sum_{j_{i1},\ldots ,j_{im},j'_{i1},\ldots ,j'_{im}} \!\!\!\bar{a}^{(\sum_{p'}
 \mu_{ip'} k'_{p'} )}_{j'_{i1},\ldots ,j'_{im}} \, a^{(\sum_p \mu_{ip}
 k_p )}_{j_{i1},\ldots ,j_{im}} \,\langle j'_{i1},\ldots ,j'_{im} |
 {\cal E}'^{\dag}_i {\cal E}_i | j_{i1},\ldots ,j_{im} \rangle \right] ~.
 \label{E:Condition_QBC_1}
\end{eqnarray}
 Substituting Eq.~(\ref{E:Condition}) into Eq.~(\ref{E:Condition_QBC_1}), we
 have
\begin{eqnarray}
 & & \langle {\bf k}_{\rm encode}' | {\cal E}'^{\dag} {\cal E} |
 {\bf k}_{\rm encode} \rangle \nonumber \\ & = & \prod_{i=1}^{+\infty} \left[
 \left\langle \left( {\textstyle \sum_p \mu_{ip} k'_p} \right)_{\rm encode}
 \left| {\cal E}'^{\dag}_i {\cal E}_i \right| \left( {\textstyle \sum_p
 \mu_{ip} k_p} \right)_{\rm encode} \right\rangle \right] \nonumber \\ & = &
 \prod_{i=1}^{+\infty} \left[ \delta_{\scriptstyle \sum_p \mu_{ip} k_p,\sum_p
 \mu_{ip} k'_p} \,\Lambda_{{\cal E}_i,{\cal E}'_i}
 \right] \label{E:Condition_QBC_2}
\end{eqnarray}
 for some constants $\Lambda_{{\cal E}_i,{\cal E}'_i}$ independent of ${\bf k}$
 and ${\bf k}'$. Since the matrix $\mu$ is invertible, $k_i = k'_i$ for all $i$
 is the unique solution of the systems of linear equations $\sum_p \mu_{ip}
 k_p = \sum_p \mu_{ip} k'_p$. Consequently,
\begin{equation}
 \langle {\bf k}_{\rm encode}' | {\cal E}'^{\dag} {\cal E} |
 {\bf k}_{\rm encode} \rangle = \delta_{{\bf k},{\bf k}'} \,\Lambda_{{\cal E},
 {\cal E}'} \label{E:Condition_QBC}
\end{equation}
 for some constant $\Lambda_{{\cal E},{\cal E}'}$ independent of ${\bf k}$ and
 ${\bf k}'$. Thus, the encoding in Eq.~(\ref{E:Encode_QBC}) is a QECC.
\hfill$\Box$
\par\bigskip\indent
 At this point, readers should realize that the above scheme can be generalized
 to construct a QCC from a QBC which maps $n$ quantum registers to $m ~(> n)$
 registers. It is also clear that the following two useful corollaries follow
 directly from Theorem~\ref{Thrm:QBC}:
\begin{cor}
 The encoding scheme given by Eq.~(\ref{E:Encode_QBC}) gives a QCC from a QBC
 provided that (1) the elements in the matrix $\mu$ are either zeros or ones;
 (2) $\mu_{ip}$ is a function of $i-p$ only; and (3) $\mu_{ip} = \mu (i-p)$
 consists of finitely many ones. \label{Cor:Zero_One}
\end{cor}
\begin{cor}
 The encoding scheme given by Eq.~(\ref{E:Encode_QBC}) gives a QCC from a QBC
 if (1) $N$ is a prime power; (2) $\mu$ is not a zero matrix; and (3)
 $\mu_{ip}$ is a function of $i-p$ only. \label{Cor:Field}
\end{cor}
\par
 Let me illustrate the above analysis by an example.
\begin{ex}
 Starting from the spin five register code in Ref.~\cite{5-reg}, one knows that
 the following QCC can correct up to one error in every five consecutive
 quantum registers:
 \begin{eqnarray}
  & & |k_1,k_2,\ldots ,k_m,\ldots \rangle \nonumber \\ & \longmapsto &
  \bigotimes_{i=1}^{+\infty} \left[ \,\frac{1}{N^{3/2}} \!\sum_{p_i,q_i,r_i =
  0}^{N-1} \!\omega_N^{(k_i + k_{i-1})(p_i + q_i + r_i) + p_i r_i} \,|p_i,q_i,
  p_i + r_i,q_i + r_i,p_i + q_i + k_i + k_{i-1} \rangle \right]
  \label{E:Five_Bit_Conv_Code}
 \end{eqnarray}
 where $k_m \in {\Bbb Z}_N$, $\omega_N$ is a primitive $N$th root of unity, and
 all additions in the state ket are modulo $N$. The rate of this code equals
 1/5. \label{Ex:Five_Reg}
\end{ex}
\par
 Although the QCC in Eq.~(\ref{E:Encode_QBC}) looks rather complicated, the
 actual encoding process can be performed readily. Because $\mu$ is invertible,
 one can reversibly map $|k_1,k_2,\ldots ,k_n,\ldots\rangle$ to $|\sum_p
 \mu_{1p} k_p,\sum_p \mu_{2p} k_p,\ldots ,\sum_p \mu_{np} k_p,\ldots\rangle$
 \cite{Pebble,SpaceTime,Reversible}. Then, one obtains the above five register
 QCC by encoding each quantum register using the procedure in
 Ref.~\cite{5-reg}.
\par
 Now, I turn to the construction of QCCs from classical convolutional codes.
 Let me first introduce two technical lemmas (which work for both QBCs and
 QCCs).
\begin{lemma}
 Suppose the QECC
 \begin{equation}
  |{\bf k}\rangle \longmapsto \sum_{j_1,j_2,\ldots} \!\! a^{({\bf k})}_{j_1,j_2
  ,\ldots} \,|j_1,j_2,\ldots \rangle \label{E:Code_Flip}
 \end{equation}
 corrects (independent) spin flip errors in certain quantum registers with $j_i
 \in {\Bbb Z}_N$. Then, the following QECC, which is obtained by discrete
 Fourier transforming every quantum register in Eq.~(\ref{E:Code_Flip}),
 \begin{eqnarray}
  |{\bf k}\rangle & \longmapsto & \sum_{j_1,j_2,\ldots ,p_1,p_2,\ldots} \!\!
  a^{({\bf k})}_{j_1,j_2,\ldots} \prod_{i=1}^{+\infty} \left(
  \frac{1}{\sqrt{N}} \,\omega_N^{j_i p_i} \right) \,|p_1,p_2,\ldots \rangle
  \label{E:Code_Phase}
 \end{eqnarray}
 corrects (independent) phase errors occurring in the same quantum registers.
 The converse is also true. \label{Lemma:Spin_Phase}
\end{lemma}
\noindent
{\it Proof:} Observe that one can freely choose a computational basis for the
 encoded quantum state. In particular, if one chooses the discrete Fourier
 transformed basis $\{ |\tilde{m}\rangle\} \equiv \{ \sum_{j=0}^{N-1}
 \omega_N^{j\, m} |j\rangle\}$ for each of the encoded quantum register, then
 the encoding in Eq.~(\ref{E:Code_Phase}) is reduced to the encoding in
 Eq.~(\ref{E:Code_Flip}). Thus, the code in Eq.~(\ref{E:Code_Phase}) handles
 spin flip errors with respected to the discrete Fourier transformed basis
 $\{ |\tilde{m}\rangle\}$. Consequently, the same code handles phase errors in
 the original $\{ |m\rangle\}$ basis.
\par
 Conversely, suppose one chooses the original $\{|m\rangle\}$ basis to encode
 a phase error correcting code. Then with respected to the $\{|\tilde{m}\rangle
 \}$ basis, it is easy to check that the same code corrects spin flip errors.
\hfill$\Box$
\begin{lemma}
 Suppose a QECC handles errors $E_1$ and $E_2$ satisfying (a) for all
 ${\cal E}_i \in E_i$ ($i=1,2$), there exists ${\cal E}'_2 \in E_2$ such that
 ${\cal E}^{\dag}_2 \circ {\cal E}_1 = {\cal E}^{\dag}_1 \circ {\cal E}'_2$;
 and (b) for ${\cal E}_i, {\cal E}'_i \in E_i$ ($i=1,2$), ${\cal E}^{\dag}_i
 \circ {\cal E}'_i \in E_i$ whenever errors ${\cal E}_i$ and ${\cal E}'_i$
 occur at the same set of quantum registers; then the QECC actually handles
 errors in $E_1 \circ E_2 \equiv \{ {\cal E}_1 \circ {\cal E}_2 : {\cal E}_1
 \in E_1, {\cal E}_2 \in E_2$ and errors ${\cal E}_1, {\cal E}_2$ occur at the
 same set of quantum registers$\}$. \label{Lemma:Compose}
\end{lemma}
\noindent
{\it Proof:} One knows from Eq.~(\ref{E:Condition}) that $\langle
 {\bf k}_{\rm encode}' | {\cal E}^{\dag}_i {\cal E}'_i | {\bf k}_{\rm encode}
 \rangle$ $= \delta_{{\bf k},{\bf k}'} \,\Lambda_{{\cal E}_i,{\cal E}'_i}$ for
 some $\Lambda_{{\cal E}_i, {\cal E}'_i}$ independent of $k$ ($i=1,2$). Also,
 Eq.~(\ref{E:Condition}) implies that the effect of an error ${\cal E}_i$ is
 simply to rigidly rotate and to contract (or expand) the encoded ket space
 independent of the state $|{\bf k}_{\rm encode}\rangle$ itself. Thus, one
 concludes that
\begin{mathletters}
\begin{equation}
 \left\langle {\bf k}'_{\rm encode} \left| \left( {\cal E}_1 + {\cal E}_2
 \right)^{\dag} \left( {\cal E}_1 + {\cal E}_2 \right) \right|
 {\bf k}_{\rm encode} \right\rangle = \delta_{{\bf k},{\bf k}'}
 \,\Gamma_{{\cal E}_1,{\cal E}_2} \label{E:Product1}
\end{equation}
\noindent and
\begin{equation}
 \left\langle {\bf k}'_{\rm encode} \left| \left( {\cal E}_1 + i {\cal E}_2
 \right)^{\dag} \left( {\cal E}_1 + i {\cal E}_2 \right) \right|
 {\bf k}_{\rm encode} \right\rangle = \delta_{{\bf k},{\bf k}'}
 \,\Gamma'_{{\cal E}_1,{\cal E}_2} \label{E:Product2}
\end{equation}
\end{mathletters}
\noindent
 for all ${\cal E}_i \in E_i$ ($i=1,2$), where $\Gamma_{{\cal E}_1,{\cal E}_2}$
 and $\Gamma'_{{\cal E}_1,{\cal E}_2}$ are independent of ${\bf k}$. By
 expanding Eqs.~(\ref{E:Product1}) and~(\ref{E:Product2}), one arrives at
\begin{equation}
 \langle {\bf k}'_{\rm encode} | {\cal E}^{\dag}_1 {\cal E}_2 |
 {\bf k}_{\rm encode} \rangle = \delta_{{\bf k},{\bf k}'} \,\Xi_{{\cal E}_1,
 {\cal E}_2} \label{E:Cross_Term}
\end{equation}
 for some $\Xi_{{\cal E}_1,{\cal E}_2}$ independent of ${\bf k}$. Finally, I
 consider errors ${\cal E}_i, {\cal E}'_i \in E_i$ ($i=1,2$) occurring at the
 same set of quantum registers, then
\begin{eqnarray}
 & & \left\langle {\bf k}'_{\rm encode} \left| \left( {\cal E}'_1 {\cal E}'_2
 \right)^{\dag} \left( {\cal E}_1 {\cal E}_2 \right) \right|
 {\bf k}_{\rm encode} \right\rangle \nonumber \\ & = & \langle
 {\bf k}'_{\rm encode} | {\cal E}'^{\dag}_2 {\cal E}'^{\dag}_1 {\cal E}_1
 {\cal E}_2 | {\bf k}_{\rm encode} \rangle \nonumber \\ & = & \langle
 {\bf k}'_{\rm encode} | {\cal E}''_1 {\cal E}''_2 | {\bf k}_{\rm encode}
 \rangle \label{E:Final_Cross}
\end{eqnarray}
 for some ${\cal E}''_i \in E_i$ ($i=1,2$). Hence from Eqs.~(\ref{E:Condition})
 and~(\ref{E:Cross_Term}), I conclude that the QECC handles errors in the set
 $E_1 \circ E_2$.
\hfill$\Box$
\par\bigskip\indent
 The next corollary follows directly from Lemma~\ref{Lemma:Compose}.
\begin{cor}
 A QECC handles general quantum error if and only if it handles both spin flip
 and phase errors in the corresponding quantum registers.
 \label{Cor:Flip_Phase}
\end{cor}
\par\bigskip\indent
 Now, I am ready to prove the following theorem regarding the construction of
 quantum codes from classical codes.
\begin{thrm}
 Suppose QECCs $C1$ and $C2$ handle phase shift and spin flip errors,
 respectively, for the same set of quantum registers. Then, pasting the two
 codes together by first encodes the quantum state using $C1$ then further
 encodes the resultant quantum state using $C2$, one obtains a QECC $C$ which
 corrects general errors in the same set of quantum registers.
 \label{Thrm:Paste}
\end{thrm}
\noindent
{\it Proof:} From Corollary~\ref{Cor:Flip_Phase}, it suffices to show that the
 new QECC $C$ corrects both spin flip and phase errors. By the construction of
 $C$, it clearly can correct spin flip errors. And using the same trick in the
 proof of Lemma~\ref{Lemma:Compose}, it is easy to check that $C$ can correct
 phase shift errors as well.
\hfill$\Box$
\par\bigskip\indent
 Readers should note that the order of pasting in Theorem~\ref{Thrm:Paste} is
 important. Reversing the order of encoding does not give a good quantum code.
 Also, proofs of Corollary~\ref{Cor:Flip_Phase} and Theorem~\ref{Thrm:Paste}
 for the case of $N = 2$ can also be found, for example, in Ref.~\cite{Cald2}.
\begin{thrm}
 Suppose $C$ is a classical (block or convolutional) code of rate $r$ that can
 correct $p$ (classical) errors for every $q$ consecutive registers. Then, $C$
 can be extended to a QECC of rate $r^2$ that can correct at least $p$ quantum
 errors for every $q^2$ consecutive quantum registers. \label{Thrm:Pasting}
\end{thrm}
\noindent
{\it Proof:} Suppose $C$ is a classical code. By mapping $m$ to $|m\rangle$ for
 all $m\in {\Bbb Z}_N$, $C$ can be converted to a quantum code for spin flip
 errors. Let $C'$ be the QECC obtained by Fourier transforming each quantum
 register of $C$. Then Lemma~\ref{Lemma:Spin_Phase} implies that $C'$ is a code
 for phase shift errors. From Theorem~\ref{Thrm:Paste}, pasting codes $C$ and
 $C'$ together will create a QECC $C''$ of rate $r^2$. Finally, one can verify
 the error correcting capability of $C''$ readily \cite{QCC}.
\hfill$\Box$
\par\bigskip
 Theorem~\ref{Thrm:Pasting} is useful to create high rate QCCs from high rate
 classical convolutional codes. Note that one of the simplest classical
 convolutional code with rate $1/2$ is given by Eq.~(\ref{E:Classical_Conv}).
 Being a non-systematic\footnote{That is, both $b_i$ and $c_i$ are not equal to
 $a_i$.} and non-catastrophic\footnote{That is, a finite number of channel
 errors does not create an infinite number of decoding errors.} code
 \cite{21-encoder}, it serves as an ideal starting point to construct good
 QCCs. First, let me write down this code in quantum mechanical form:
\begin{lemma}
 The QCC
 \begin{equation}
  |k_1,k_2,\ldots\rangle \longmapsto \bigotimes_{i=1}^{+\infty} |k_i + k_{i-2},
  k_i + k_{i-1} + k_{i-2} \rangle \label{E:21_QSpin}
 \end{equation}
 for all $k_i \in {\Bbb Z}_N$, where all additions in the state ket are modulo
 $N$, can correct up to one spin flip error for every four consecutive quantum
 registers. \label{Lemma:21-encoder}
\end{lemma}
\noindent
{\it Proof:} Using notations as in the proof of Theorem~\ref{Thrm:QBC}, I
 consider $\langle {\bf k}'_{\rm encode} | {\cal E}'^{\dag} {\cal E} |
 {\bf k}_{\rm encode} \rangle$. Clearly, the worst case happens when errors
 ${\cal E}$ and ${\cal E}'$ occur at different quantum registers. And in this
 case, Eq.~(\ref{E:21_QSpin}) implies that exactly two of the following four
 equations hold:
 \begin{equation}
  \left\{ \begin{array}{rcl} k_{2i} + k_{2i-2} & = & k'_{2i} + k'_{2i-2} \\
  k_{2i} + k_{2i-1} + k_{2i-2} & = & k'_{2i} + k'_{2i-1} + k'_{2i-2} \\
  k_{2i+1} + k_{2i-1} & = & k'_{2i+1} + k'_{2i-1} \\ k_{2i+1} + k_{2i} +
  k_{2i-1} & = & k'_{2i+1} + k'_{2i} + k'_{2i-1} \end{array} \right.
  \label{E:Systems_1}
 \end{equation}
 for all $i$. One may regard $k_i$s as unknowns and $k'_i$s as arbitrary but
 fixed constants. Then, by straight forward computation, one can show that
 picking {\em any} two equations out of Eq.~(\ref{E:Systems_1}) for each $i$
 will form an invertible system with the unique solution $k_i = k'_i$ for all
 $i$. Thus, $\langle {\bf k}'_{\rm encode} | {\cal E}'^{\dag} {\cal E} |
 {\bf k}_{\rm encode} \rangle = \delta_{{\bf k},{\bf k}'} \,\delta_{{\cal E},
 {\cal E}'}$ and hence this lemma is proved.
\hfill$\Box$
\begin{ex}
 Theorem~\ref{Thrm:Pasting} and Lemma~\ref{Lemma:21-encoder} imply that the
 following QCC of rate 1/4:
 \begin{eqnarray}
  & & |k_1,k_2,\ldots \rangle \longmapsto |{\bf k}_{\rm encode} \rangle \equiv
  \nonumber \\ & & \bigotimes_{i=1}^{+\infty} \left[ \sum_{p_1,q_1,\ldots}
  \!\frac{1}{N} \,\omega_N^{(k_i + k_{i-2}) p_i + (k_i + k_{i-1} + k_{i-2})
  q_i} \,|p_i + p_{i-1}, p_i + p_{i-1} + q_{i-1},q_i + q_{i-1},q_i + q_{i-1} +
  p_i \rangle \right] \label{E:14QCC}
 \end{eqnarray}
 for all $k_i \in {\Bbb Z}_N$, where all additions in the state ket are modulo
 $N$ can correct at least one error for every 16 consecutive quantum registers.
 But, in fact, this code is powerful enough to correct one error for every
 eight consecutive quantum registers (see also Ref.~\cite{QCC}).
 \label{Ex:4-reg-code}
\end{ex}
\noindent
{\it Proof:} Let ${\cal E}$ and ${\cal E}'$ be two quantum errors affecting at
 most one quantum register per every eight consecutive ones. By considering
 $\langle {\bf k}'_{\rm encode} | {\cal E}'^{\dag} {\cal E} |
 {\bf k}_{\rm encode} \rangle$, I know that at least six of the following eight
 equations hold:
\begin{equation}
 \left\{ \begin{array}{rcl} p_{2i-1} + p_{2i-2} & = & p'_{2i-1} + p'_{2i-2} \\
 p_{2i-1} + p_{2i-2} + q_{2i-2} & = & p'_{2i-1} + p'_{2i-2} + q'_{2i-2} \\
 q_{2i-1} + q_{2i-2} & = & q'_{2i-1} + q'_{2i-2} \\ q_{2i-1} + q_{2i-2} +
 p_{2i-1} & = & q'_{2i-1} + q'_{2i-2} + p'_{2i-1} \\ p_{2i} + p_{2i-1} & = &
 p'_{2i} + p'_{2i-1} \\ p_{2i} + p_{2i-1} + q_{2i-1} & = & p'_{2i} + p'_{2i-1}
 + q'_{2i-1} \\ q_{2i} + q_{2i-1} & = & q'_{2i} + q'_{2i-1} \\ q_{2i} +
 q_{2i-1} + p_{2i} & = & q'_{2i} + q'_{2i-1} + p'_{2i} \end{array} \right.
 \label{E:Systems_2}
\end{equation}
 for all $i\in {\Bbb Z}^{+}$. Let me regard $p_i$ and $q_i$ as unknowns; and
 $p'_i$ and $q'_i$ as arbitrary but fixed constants. Then, it is straight
 forward to show that choosing {\em any} six equations in
 Eq.~(\ref{E:Systems_2}) for each $i\in {\Bbb Z}^{+}$ would result in a
 consistent system having a unique solution of $p_i = p'_i$ and $q_i = q'_i$
 for all $i\in {\Bbb Z}^{+}$. Consequently,
\begin{eqnarray}
 & & \langle {\bf k}'_{\rm encode} | {\cal E}'^{\dag} {\cal E} |
 {\bf k}_{\rm encode} \rangle \nonumber \\ & = & \sum_{p_1,q_1,p_2,q_2,\ldots}
 \left\{ \prod_{i=1}^{+\infty} \left[ \,\omega_N^{\sum_{j=2i-1}^{2i} p_j (k_j +
 k_{j-2} - k'_j - k'_{j-2}) + q_j (k_j + k_{j-1} + k_{j-2} - k'_j - k'_{j-1} -
 k'_{j-2})} \,\langle f_i | {\cal E}'^{\dag}_i | f_i \rangle \,\langle g_i |
 {\cal E} | g_i \rangle \right] \right\} \label{E:Systems_3}
\end{eqnarray}
 for some linearly independent functions $f_i (p_1,q_1,p_2,q_2,\ldots )$ and
 $g_i (p_1,q_1,p_2,q_2,\ldots )$.
\par
 Now, I consider a basis $\{ h_i (p_1,q_1,p_2,q_2,\ldots ) \}$ for the
 orthogonal complement of the span of $\{ f_i, g_i \}_{i\in {\Bbb Z}^{+}}$. By
 summing over all $h_i$s while keeping $f_i$s and $g_i$s constant in
 Eq.~(\ref{E:Systems_3}), one ends up with the constraints that $k_i = k'_i$
 for all $i\in {\Bbb Z}^{+}$. Thus,
\begin{eqnarray}
 & & \langle {\bf k}'_{\rm encode} | {\cal E}'^{\dag} {\cal E} |
 {\bf k}_{\rm encode} \rangle \nonumber \\ & = & \delta_{{\bf k},{\bf k}'}
 \sum_{p_1,q_1,p_2,q_2,\ldots } \left[ \prod_{i=1}^{+\infty} \left( \langle f_i
 (p_1,q_1,\ldots ) | {\cal E}'^{\dag} | f_i (p_1,q_1,\ldots ) \rangle \,\langle
 g_i (p_1,q_1,\ldots ) | {\cal E} | g_i (p_1,q_1,\ldots ) \rangle \right)
 \right] ~. \label{E:Systems_4}
\end{eqnarray}
 Hence, Eq.~(\ref{E:14QCC}) corrects up to one quantum error per every eight
 consecutive quantum registers.
\hfill$\Box$
\par\bigskip
 The above rate 1/4 QCC is constructed from a classical convolutional code of
 rate 1/2. One may further boost up the code performance by converting other
 efficient classical convolutional codes (such as various $k/(k+1)$-rate codes
 in Ref.~\cite{k+1codes}) into QCCs. On the other hand, it is impossible to
 construct a four quantum register QBC that can correct one quantum error
 \cite{5-reg,Cond1}. With modification, the same argument can be used to show
 that no QCC can correct one error for every four consecutive quantum registers
 \cite{QCC}. It is instructive to compare the performances of QBCs and QCCs in
 other situations.
\par
 In addition, in order use QCCs in quantum computation, one must investigate
 the possibility of fault tolerant computation on them. Moreover, it would be
 ideal if the fault tolerant implementation of single and two-quantum register
 operations must involve only a finite number of quantum registers in the QCC.
 While a general QCC may not admit a finite fault tolerant implementation, many
 QCCs with finite memories\footnote{That is, codes with encoding schemes which
 depend on a finite number of quantum registers in $|{\bf k}\rangle$.} can be
 manipulated fault tolerantly.
\begin{ex}
 By subtracting those quantum registers containing $p_i$, $p_{i+2}$, $q_i$,
 $q_{i+1}$ and $q_{i+2}$ by one in Eq.~(\ref{E:14QCC}), one ends up with
 changing $|k_1,k_2,\ldots,k_i,\ldots_{\rm encode}\rangle$ to $|k_1,k_2,\ldots,
 k_{i-1},k_i+1,k_{i+1},\ldots_{\rm encode}\rangle$. Clearly, the above
 operation is fault tolerant and involves only a finite number of quantum
 registers. Fault tolerant implementation of single register phase shift can be
 obtained in a similar way. Further results on fault tolerant implementation on
 QCCs will be reported elsewhere \cite{Further}. \label{Ex:Fault_Tol}
\end{ex}
 Finally, decoding a classical convolutional code can be quite involved
 \cite{Conv_Decode}. So, it is worthwhile to investigate the efficiency of
 decoding a QCC. I shall report them in future works \cite{Further}.
\acknowledgments
 I would like to thank T.~M. Ko for introducing me the subject of convolutional
 codes. I would also like to thank Debbie Leung, H.-K. Lo and Eric Rains for
 their useful discussions. This work is supported by the Hong Kong Government
 RGC grant HKU~7095/97P.


\begin{references}
\bibitem{k+1codes} A list of good classical convolutional codes can be found,
 for example, in appendix~A of Ref.~\cite{Convolution2}.
\end{references}
\end{document}